\input{aipcheck}
\def\ltsima{$\; \buildrel < \over \sim \;$}
\def\simlt{\lower.5ex\hbox{\ltsima}}
\def\gtsima{$\; \buildrel > \over \sim \;$}
\def\simgt{\lower.5ex\hbox{\gtsima}}

\documentclass[final,sort]{aipproc}

\layoutstyle{6x9}
\usepackage{psfig}

\begin{document}

\title{X-ray obscuration in local Universe AGN}


\author{Matteo Guainazzi}{
  address={European Space Astronomy Center, Apartado 50727, E-28080 Madrid, Spain}
}

\begin{abstract}
I review the constraints that X-ray observations impose
on the physical properties and the geometrical distribution of cold
absorbing gas in nearby obscured Active Galactic Nuclei (AGN),
as well as their implications
for AGN structure models.
\end{abstract}

\maketitle


\section{Introduction}

It has been known since the dawn of X-ray spectroscopy
that the high-energy emission of type~2 Active Galactic Nuclei (AGN)
is obscured by large column densities of cold and neutral gas
\citep{turner89,awaki91}. This discovery nicely matches the predictions
of the standard Seyfert unification scenario \citep{antonucci85,antonucci93},
which ascribes the ``type~1'' versus ``type~2'' dichotomy to our line-of-sight
orientation with respect to an azimuthally symmetric dust molecular
``torus'', preventing the direct view of the central engine and of
the Broad Line Regions (BLRs) in the latter.

After almost twenty years of Seyfert~2 galaxies
X-ray observations: which constraints can we
derive on the physical properties and on the geometrical
distribution of the X-ray obscuring matter? Which relations exist between the
X-ray absorbing gas and material responsible for obscuration or reddening
at other wavelengths? Which are the implications of these results for the
AGN structure models? I will address these question in this contribution.

A bit of terminology before starting: X-ray obscured AGN can be classified
into ``Compton-thin'' or ``-thick'' depending on the column density, $N_H$,
covering the nuclear emission. In objects belonging to the
latter class $N_H \simgt \sigma_t^{-1} \sim 1.6 \times 10^{24}$~cm$^{-2}$.
The direct nuclear emission of Compton-thick Seyfert~2s is
totally suppressed in the 0.1--10~keV energy band, where most of
sensitive X-ray observations have been performed so far.
Even in this band, however,
Compton-thick Seyfert~2 galaxies
are not totally silent: the flat continuum
measured in most of them, together with the large Equivalent
Width ($EW > 1$~keV) iron K$_{\alpha}$ fluorescent line, indicates a
dominant contribution by Compton-reflection off optically-thick matter,
most likely illuminated by the intrinsic AGN continuum through an unobscured
optical path. Traditionally, X-ray spectra of Compton-thick object
are referred to as ``reflection-dominated'' (and those of Compton-thin
objects as ``transmission-dominated''). Although there are good reasons
to believe that this association does not always hold
\citep{matt03a,guainazzi05a}
we will maintain it throughout this paper for simplicity.
Finally, some of the experimental evidence presented in this
paper refer to optically-thick X-ray reprocessing rather than to
X-ray absorption. Whenever pertinent, I will make explicit the physical
reasons, allowing us to extend to the latter the conclusions
derived for the former.

\section{Properties of optically-thick gas in the AGN environment from
X-ray observations}

\subsection{Column density distribution}

Several independent studies of optically-selected Seyfert~2s
samples suggest that the distribution of
column densities in the local Universe is flat
in logarithmic scale across the whole range between 10$^{22}$ and
10$^{25}$~cm$^{-2}$
\citep{risaliti99,guainazzi05b}\footnote{On the
other hand, radio-loud
samples seem to be missing highly obscured AGN \citep{grandi06}.}.
They are confirmed by the results of a complete, volume-limited
sample of AGN observed by XMM-Newton \citep{cappi06}.
On the basis of a O[{\sc iii}] selected sub-sample of the
RXTE All-Sky Survey \citep{revnivtsev04}, Heckman et al. \cite{heckman05}
estimate that the fraction of Compton-thick AGN
over the total number of Seyfert~2 galaxies
in the local Universe is $36 \pm 14 \%$.
 The existence of
``elusive'' AGN \citep{maiolino03}, where
optical signatures of nuclear activity
are missing, implies that the above estimate could be actually
only a lower limit to the true number.

Such a large fraction of highly-obscured AGN should be
present at cosmological distances as well. The $N_H$ distribution
can be constrained by comparing the spectrum of
the Cosmic X-ray Background (CXB) with models producing it
through the superposition of unresolved AGN \citep{setti89,comastri95}.
Indeed, the intrinsic column density distribution in large X-ray surveys
is consistent with being flat \citep{lafranca05}. 
Nonetheless, recent
results from a survey of more than 127 AGN detected by INTEGRAL IBIS/ISGRI
above 20~keV \citep{sazonov06} unveil a fraction
of Compton-thick AGN ($\sim$10\%)
significantly lower than predicted by
CXB models. However, as the same authors point out in their paper,
this result has to be taken with caution, because
the X-ray emission of ``very Compton-thick
objects'' ($N_H \simgt 10^{25-26}$~cm$^{-2}$)
is expected to be significantly suppressed by
Compton scattering even in the INTEGRAL
energy band.

\subsection{Location}

The measurement of the soft photoelectric spectral
cut-off allows us to measure
the column density integrated along the whole line of sight to the
AGN. Hence, X-rays can provide constraints on the distribution and location
of optically thick gas only
from direct high-resolution imaging observations,
and from variability studies.

In the Circinus~Galaxy \citep{sambruna01}
images in the iron K$_{\alpha}$ fluorescent emission line
band, likely to be dominated by optically-thick reprocessing of the
nuclear continuum, are unresolved at the {\it Chandra} spatial resolution
($\sim$0.3$^{{\prime}{\prime}}$). This constrains the size of the
line-emitting region to be $\simlt$8~pc. 
Interestingly enough, in NGC~1068 the hard X-ray emission,
including the iron line region, is seen extending $\sim$2~kpc \citep{young01},
with a tentative detection of iron K$_{\alpha}$ fluorescent line
extension up to $\sim$5.5~kpc. This evidence is in contradiction
with the compact size of the torus as derived from the interferometric
measurements in this object \citep{jaffe04}. In more distant
Compton-thick AGN the constraints drawn from direct
iron imaging are looser. However, indirect constraints
on the compactness of the reflector can be drawn from the lack of
photoelectric ``shadows''
on AGN-photoionized gas extending on
scales as large as 0.1-1~kpc \citep{bianchi06}. 

Evidence for compact cold gas in the nuclear environment
comes also from
the detection of rapid variability of the X-ray absorber.
A breakthrough discovery in this field was the detection of
one order of magnitude variation of the absorber column density
between two RXTE observations of NGC~4388 taken 4 hours
apart \citep{elvis04}. If
explained by occultation of a gas cloud ($N_H = 3 \times 10^{22}$)
in Keplerian motion around the super-massive black hole, the
inferred cloud location ($\sim$10$^3$ Schwarzschild radii) is
typical of the BLR scale rather than of the standard ``torus''
envisaged by the Seyfert unification scenario. The same conclusions
were drawn by Puccetti et al. (2004)
\cite{puccetti04} from a similar episode detected
in NGC~4151. An even more extreme
episode of variability has been recently detected during an intense
monitoring campaign of the Compton-thick Seyfert~2 galaxy NGC~1365.
This AGN has been shown to exhibit changes from a ``reflection-''
to a ``transmission-dominated'' status in less
than two days (See Elvis et al., this volume). All these pieces
of observational evidence suggest that, at least in some
objects, the gas responsible for the X-ray obscuration is
close to the central engine. Risaliti et al. (2002) speculate that it could
be associated with accretion disk dust-free outflows, in the
framework of the model of AGN structure proposed by
Elvis (2000).

On the other hand, observational
evidence exists, which associate the Compton-thin
X-ray absorber with matter located on scales much larger than the
nuclear environment. First of all, good quality spectra
of several nearby AGN exhibit simultaneous signatures of
obscuration and/or reprocessing by Compton-thick matter and of
absorption by Compton-thin matter (NGC~424, \citep{matt03b};
NGC~5506, \citep{bianchi03}; NGC~7582, Piconcelli et al., in
preparation).The spectral transitions observed in
``changing-look'' Seyfert~2 galaxies \citep{matt03a,guainazzi05a}
can be explained as a change in the optical path along which we are
looking at the AGN emission, due to extreme variability of the
AGN intrinsic output. In either state, the optical path intercepts
a compact Compton-thick absorber, or a Compton-thin absorber external
to the former, respectively
(see as well Guainazzi et al. 2004 \cite{guainazzi04}
for a possible case
of a Compton-thin absorber covering the spectrum
produced by Compton-thick reprocessing). Finally, a study
of the X-ray absorption properties of a Seyfert~2 sample selected
according to nuclear dust morphology
on $\simgt$100~pc scale \citep{malkan98} led
Guainazzi et al. (2005) \cite{guainazzi05b}
to conclude that Compton-thin Seyfert~2s are
preferentially located in ``dusty'' nuclear environments, whereas
Compton-thick Seyfert~2 galaxies are equally distributed between
nuclear ``dusty'' and ``not-dusty'' host galaxies. These pieces
of evidence suggest that X-ray Compton-thick and -thin
absorbers are physically decoupled. The former is
probably very compact
(typical scales $\simlt$1~pc), the latter is likely to be associated
to host galaxy dust lanes on scales $\simgt 100$~pc. The possible
role of host galaxy dust in shaping the AGN classification had
already been suggested by Maiolino \& Rieke (1995) and Malkan et
al (1998) \cite{maiolino95,malkan98}.

\subsection{Geometry and covering fraction}

It is difficult to get direct X-ray observational constraints on the
geometry of the X-ray absorber and/or reflector. One of the most promising
methods is the detection of the iron K$_{\alpha}$ fluorescent line
Compton Shoulder in objects, where the column density of the gas
covering the nuclear emission is known. The relative intensity of
the Compton Shoulder with respect to the
K$_{\alpha}$ is a function of the reflector
column density and of the line-of-sight inclination \citep{sunyaev96,matt02}.
The measurement of the Compton Shoulder in the {\it Chandra} HETG spectrum
of the Circinus Galaxy \citep{bianchi01} led to the conclusion that
the column density of the absorber and of the reflector are the same.
If one associate the former (latter) with the closer (farther) side
of the same gaseous system, this must have a toroidal symmetry as the
``torus'' envisaged by the Seyfert unification scenarios.

\begin{figure}
 \hbox{
  \includegraphics[height=.35\textheight]{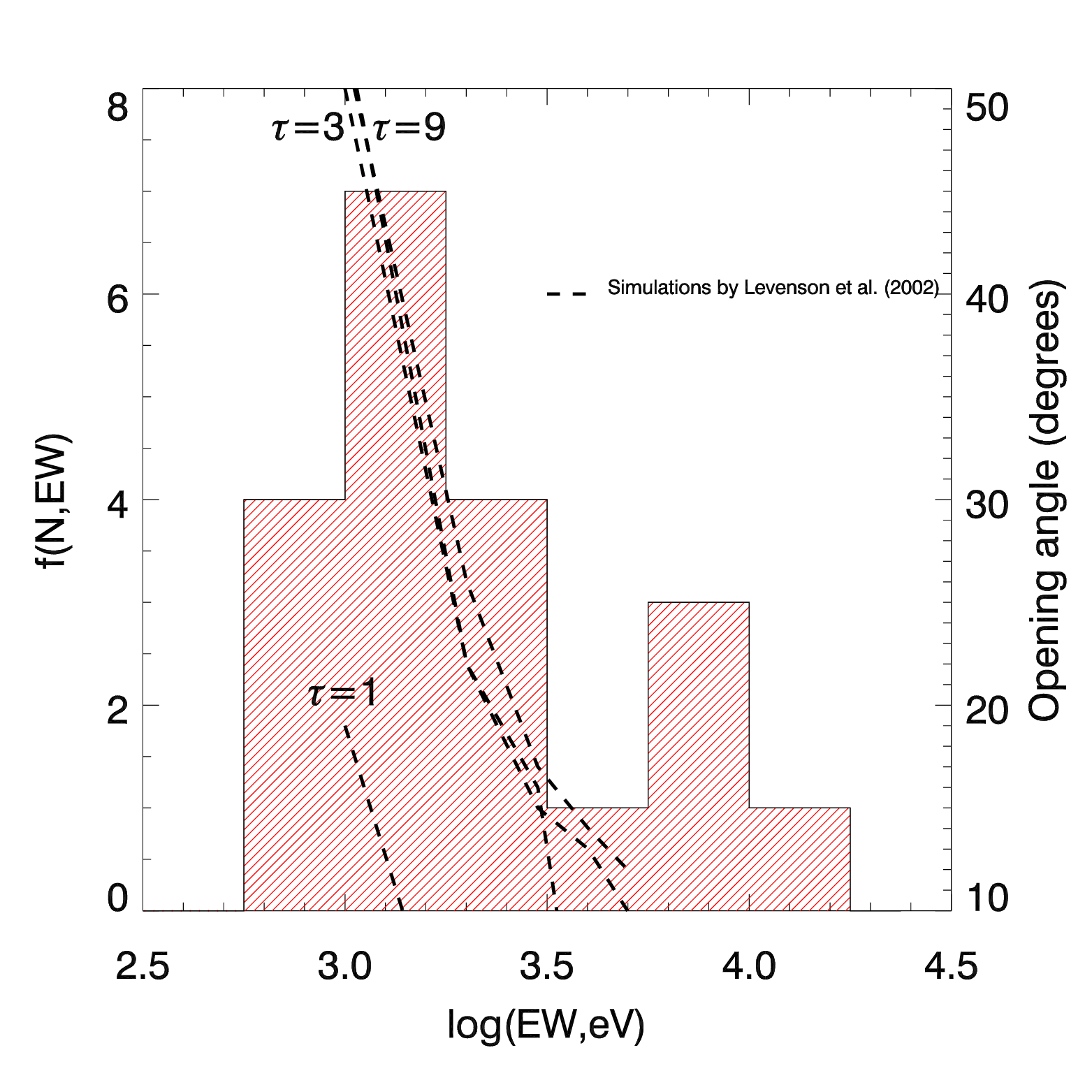}
  \hspace{0.5cm}
  \includegraphics[height=.325\textheight]{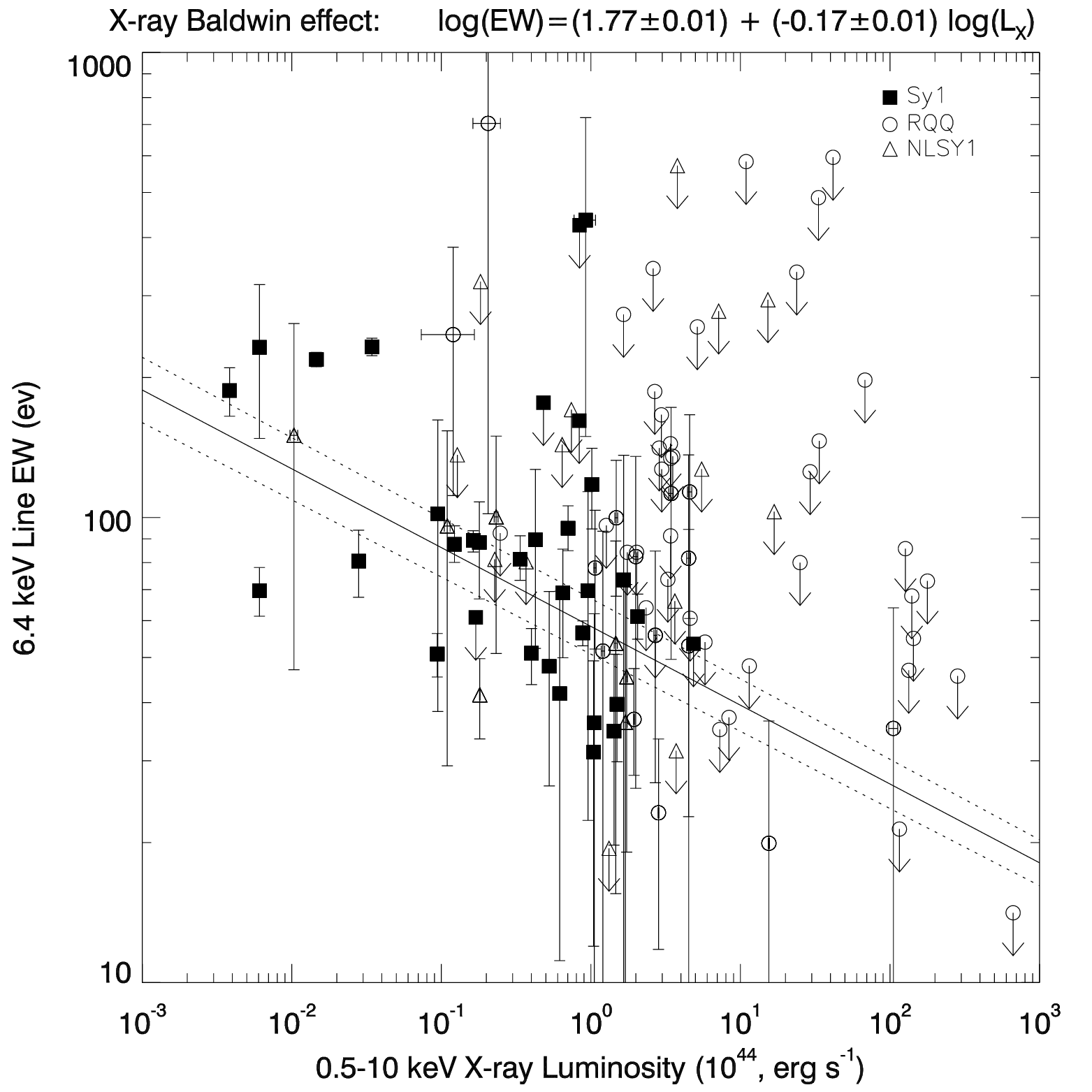}
  }
  \caption{{\it Left panel}: the {\it hatched histogram} represents the
  iron K$_{\alpha}$ fluorescent line EW distribution
  in an optically selected sample
  of Compton-thick Seyfert~2 galaxies observed by XMM-Newton
  \citep{guainazzi05b}. The {\it dashed lines} represent the opening angle
  of optically thick gas in a torus-like structure required to produce
  a given EW for different values of the Thompson optical depths
  \citep{krolik94,levenson02,levenson05}. {\it Right panel}:
  anti-correlation between the iron K$_{\alpha}$ fluorescent line EW and
  the X-ray luminosity (the''X-ray Baldwin effect'' or
  ``Iwasawa effect'')
  in a large sample of unobscured AGN observed by XMM-Newton (Bianchi et al.,
  in preparation).}
  \label{fig1}
\end{figure}

Regardless of the detailed geometry, the covering factor of the reflector
must be large to account for the relative intensity of the Compton-reflection
component in reflection-dominated spectra (see, {\it e.g.}, the
discussion in Risaliti et al. (2005) \cite{risaliti05}),
as well as for the iron K$_{\alpha}$ fluorescent line
EWs measured in Compton-thick
objects. In the {\it left panel} of Fig.~\ref{fig1} we compare the
distribution of the EWs in an optically-defined sample of nearby ($z \le
0.035$) Compton-thick Seyfert~2 galaxies observed by XMM-Newton
\citep{guainazzi05b} with the predictions of Monte-Carlo simulations
\citep{krolik94,levenson02,levenson05}. Even for extreme Thompson
optical depths, EWs larger than 1~keV require opening angles lower than
30$^{\circ}$, corresponding to 
a solid angle covering $\simgt$80\% of the sky as seen by the
central engine.

The covering fraction of the circumnuclear obscuring/absorbing gas
is probably not a cosmological invariant. There are at least two independent
pieces of experimental evidence supporting this conclusion. The
existence of the so-called
``Iwasawa effect'' (an anti-correlation between the EW of the iron
K$_{\alpha}$ fluorescent line and the X-ray luminosity\footnote{In this
context, we discuss only of the Iwasawa effect applied to the narrow
component of the iron K$_{\alpha}$ fluorescent line. A discussion
on a possible similar effect affecting the relativistically broadened
component of the same emission line can be found in Guainazzi et al.
2006 \cite{guainazzi06}}, originally
discovered by Iwasawa \& Taniguchi (1993) \cite{iwasawa93})
seems now to be definitively
demonstrated (Bianchi et al., in preparation; see the {\it right panel}
of Fig.~\ref{fig1}), after some controversy
on possible sample selection effects \citep{page04,jimenezbailon05}.
Moreover, the analysis of different X-ray surveys suggests that
the fraction of AGN obscured by a column density larger than
10$^{22}$~cm$^{-2}$ decreases with the 2--10~keV luminosity. Both the
above pieces of evidence can be
simultaneously explained by the ``receding torus'' model,
whereby the covering fraction of the obscuring/reflecting gas decreases
with luminosity due to a flattening of the dust distribution when
the radiation pressure is stronger \citep{konigl94}. Alternatively,
low-luminosity AGN could more likely contain dusty walls 
supported by radiation pressure from a circumnuclear starburst
\cite{ohsuga01}.
The possible role of a luminosity-dependent ionization parameter was
originally discussed by Mushotzky \& Ferland (1984) \cite{mushotzky84},
and, more recently,
by Nayakshin (2000a, 2000b) \cite{nayakshin00a,nayakshin00b}
in the context of a disc origin for the
narrow iron K$_{\alpha}$ fluorescent iron line.

\subsection{Metallicity}

The metallicity of the reflector can be
in principle determined through measurements of
the iron K$_{\alpha}$ photoionization edge at $\simgt$7~keV in
reflection-dominated spectra. However,
this measurement is very challenging, due to the drop in
the effective area of CCD detectors above 6~keV. So far, this
measurement has been possible only on NGC~1068 \citep{matt04} and the
Circinus~Galaxy \citep{molendi03}. In both cases, an iron overabundance
by a factor 1.5-2.5 is required. This is in good agreement with independent
pieces of evidence, which suggest very high-metallicity for the gas in
the nuclear environment \citep{groves06,nagao06}.

\section{Implications for AGN structure models}

One of the immediate consequences of the standard Seyfert unification
scenarios is that type~1 AGN should be unaffected by obscuration from the
gas, which prevents the direct view of the BLRs. This
prediction can now be tested on a sound statistical basis, thanks to
large samples of AGN discovered in {\it Chandra} and
XMM-Newton surveys. This basic prediction is verified. There
are still a number of exceptions: the fraction of
X-ray obscured type~1 AGN ranges
between 10\% and 30\% depending on the sample
\citep{caccianiga04,mateos05,tozzi06,cappi06}. The column densities
associated to these obscured type~1 AGN are generally lower than
10$^{22}$~cm$^{-2}$. They can be easily accounted for by
matter (dust lanes, molecular clouds, etc.) in the host galaxy. As such, this
evidence does not pose any substantial problem for the viability of the
unified scenarios.

More challenging has been the discovery of a fair number of Seyfert~2
galaxies, which do not show any X-ray obscuration \citep{pappa01,panessa02}.
Wolter at al. (2005) \cite{wolter05}
even report the discover of three type~2 QSO objects
(X-ray luminosity $\sim$10$^{44}$~erg~s$^{-1}$), whose X-ray spectrum is
fully unobscured. Variability is a possible explanation, as the optical
observations on which the AGN classification is based are almost never
simultaneous with the X-ray observations. However, this explanation can
be ruled
out in the only case, for which simultaneous observations in the two bands
were purposely performed (Mkn~993, \citep{corral05}). Alternatively, the
mismatch between the optical and
the X-ray classification might point to the existence
of ``pure'' type~2 objects, where the BLRs are absent. An
interesting explanation has been proposed by Nicastro (2000) and Nicastro
et al. (2003) \cite{nicastro00,nicastro03},
who suggested that BLRs are formed by accretion
disk instabilities occurring in proximity of the radius where the disk
changes from a gas pressure to a radiation pressure dominated regime. In this
context, the existence of BLRs requires a minimum critical accretion rate:
below such a threshold, the transition radius becomes smaller than the
innermost stable orbit, and BLRs cannot form.

\begin{figure}
 \hbox{
  \psfig{file=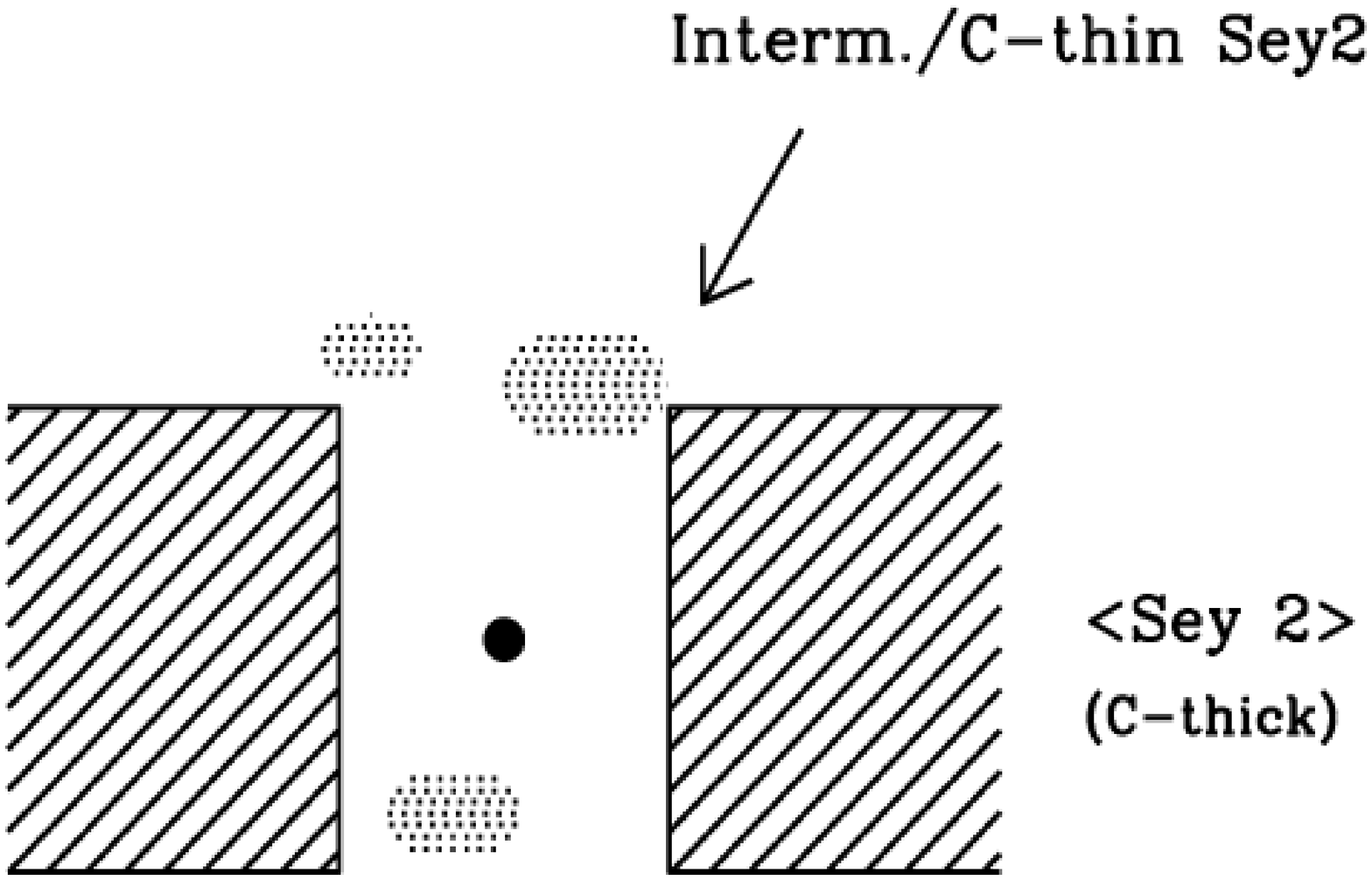,width=3.0in,height=2.5in,angle=-90}
  \hspace{0.5cm}
  \psfig{file=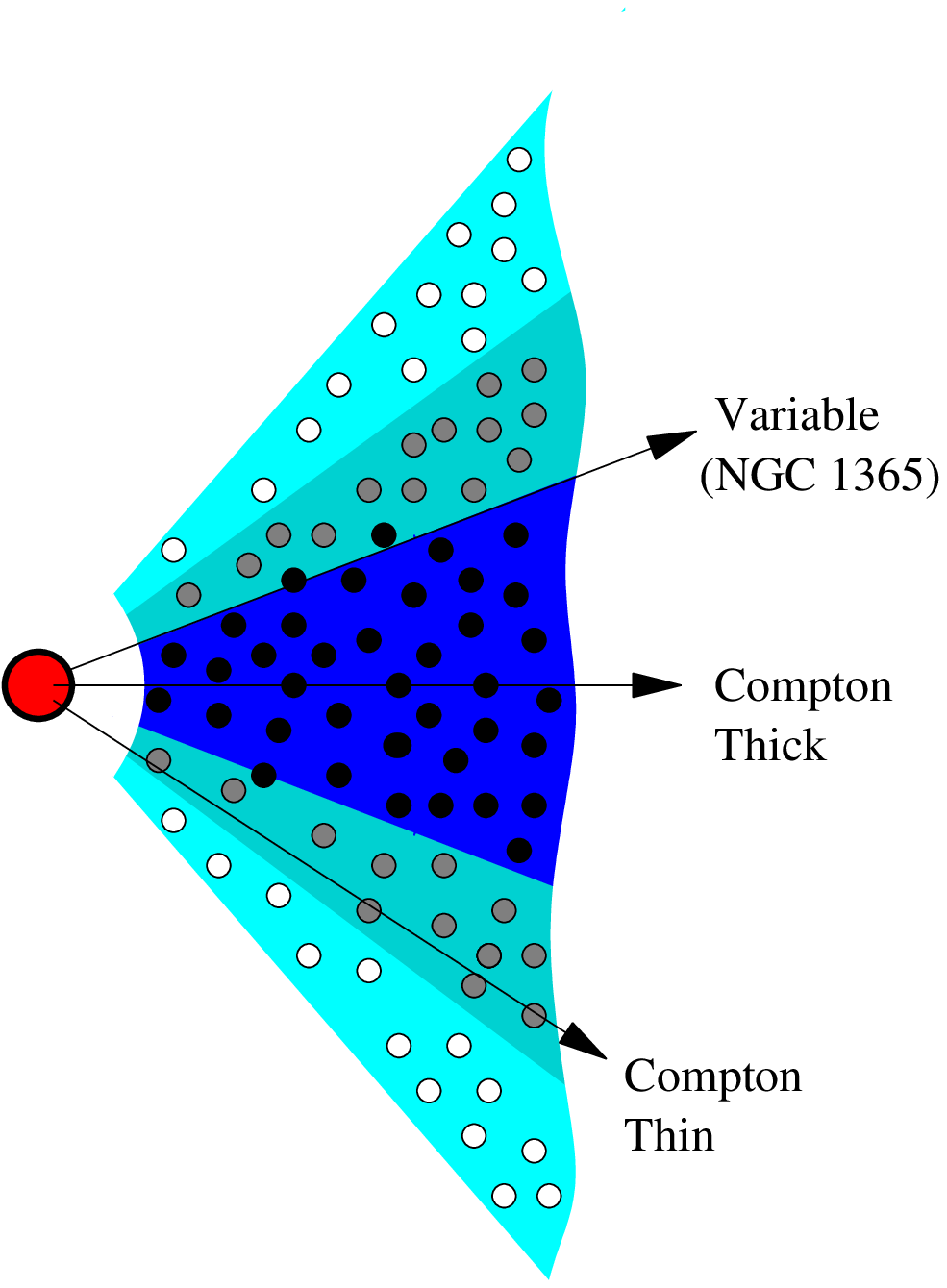,width=2.0in,height=2.5in}
  }
  \caption{{\it Left panel}: Scheme of the unification model for
   obscured AGN after Matt (2000) \cite{matt00a};
   {\it hatched rectangles}: Compton-thick
   gas; {\it dotted ellipses}: large scale Compton-thin material.
   The figure is not in scale. {\it Right panel}: scheme of the
   X-ray obscuring system for NGC~1356, after Risaliti et al.
   (2005) \cite{risaliti05}}
  \label{fig2}
\end{figure}

Fig.~\ref{fig2} shows two simplified schemes of the X-ray absorber
geometry, as proposed by Matt (2000) and Risaliti et al. (2005),
respectively \cite{matt00a,risaliti05}.
Both schemes
envisage a physical decoupling between Compton-thick and Compton-thin
obscuration. In the former scheme,
Compton-thick obscuration is primarily due to
compact optically-thick matter in the nuclear environment; the most
straightforward identification for this absorbing system (buy not the
only one!) is the ``torus'' envisaged by the Seyfert unification scenarios.
Compton-thin obscuration should be instead due to matter located on
much larger scales, possibly associated to the host galaxy. A
viable dynamical
model in this framework has been recently presented by Lamastra et al.
(2006) \cite{lamastra06}.
In the latter scheme, the X-ray absorbing system is due to a disk outflow,
whose column density decrease with azimuth (see Matt et al. (2000)
\cite{matt00b} for
tentative evidence of a dependence of the X-ray column density on the
disk-torus system inclination angle). Notwithstanding possible further
complications due to the contribution of region of intense star
formation to the X-ray obscuration \citep{ohsuga01}, or to the interplay
between optically-thick matter and highly-ionized outflows \citep{blustin05,
kinkhabwala02}, each of the above scenarios explains some of the evidence
presented in this paper, but none explains
all of them simultaneously. It is therefore
likely that both of them are telling us part of the truth. The true
geometry and distribution of the obscuring matter is
probably rather complex. Different components may correspond to different
spatial scales. Some of these components are affected by dynamical
or magneto-hydrodynamical instabilities, which imprint their signature
on the variability of the X-ray obscuration detected in some objects
\citep{risaliti02}. On the other hand,
any physically motivated models for the ``dusty torus''
require an extended structure, where
the ratio between the outer and the inner radius
is $\simeq$10--100 \citep{efstathiou95,granato97,nenkova02}. 

\begin{theacknowledgments}
I warmly thank Dr. S.~Bianchi for providing me with the
right panel of Fig.~\ref{fig1} prior to publication.
This paper reflects intense discussions with a number of collaborators:
S.~Bianchi, A.C.~Fabian, K.~Iwasawa, G.~Matt, G.C.~Perola, E.~Piconcelli,
P.~Rodriguez-Pascual.
This paper is partly based on observations obtained with XMM-Newton, an ESA
science mission with instruments and contributions directly funded by
ESA Member States and the USA (NASA).
\end{theacknowledgments}

\bibliographystyle{aipproc}   

\end{document}